\pdfoutput=1

\documentclass[10pt]{article}
\usepackage[letterpaper]{geometry}
\usepackage{hicss}
\usepackage{times}
\usepackage[none]{hyphenat}
\usepackage{url}
\usepackage{latexsym}
\usepackage{minted}
\usepackage{indentfirst}
\usepackage{graphicx}
\graphicspath{{images/}}
\usepackage[
    style=apa,
  ]{biblatex}

\usepackage{tabularx}
\usepackage{array,ragged2e} 
\usepackage{hyperref} 

\usepackage{enumitem}

\newcolumntype{P}[1]{>{\RaggedRight\arraybackslash}p{#1}}

 \usepackage[normalem]{ulem}
 \useunder{\uline}{\ul}{}

 \usepackage{lipsum} %
\usepackage{tabularx} %
\usepackage{ragged2e} %

\usepackage{quoting} %
\usepackage{adjustbox}

\usepackage{multirow}
\usepackage{booktabs}
\usepackage{tabularx}
\usepackage{xcolor}
\graphicspath{ {./images/} }

\usepackage[most]{tcolorbox}

\newtcolorbox{illustrativecasebox}{
colback=gray!5,
colframe=gray!60,
boxrule=0.6pt,
arc=2pt,
left=6pt,
right=6pt,
top=6pt,
bottom=6pt
}

\title{Barriers to GenAI Integration During Large-Scale Agile Transformation: An Exploratory Case Study}

\author{
\begin{tabular}[t]{c}
Nils Brede Moe\\
SINTEF\\
\underline{nils.b.moe@sintef.no}
\end{tabular}
\hspace{0.21cm}
\begin{tabular}[t]{c}
Marthe Berntzen \\
University of Oslo\\
\underline{marthenb@ifi.uio.no}
\end{tabular}
\hspace{0.21cm}
\begin{tabular}[t]{c}
Astri Barbala\\
SINTEF \\
\underline{astri.barbala@sintef.no}
\end{tabular}
\hspace{0.21cm}
\begin{tabular}[t]{c}
Viktoria Stray\\
University of Oslo, SINTEF \\
\underline{stray@ifi.uio.no}
\end{tabular}
}

\date{}

\begin{document}
\maketitle

\begin{abstract}
Organizations undergoing large-scale agile transformations are increasingly introducing generative AI (GenAI) tools into everyday work. While GenAI promises to accelerate software development, less is known about how organizations integrate GenAI into existing large-scale agile ways of working and how it interacts with tensions related to autonomy, alignment, learning, and coordination. We examine barriers to GenAI integration through an exploratory case study of a large Nordic media organization undergoing a large-scale agile transformation. The findings reveal team-level barriers, including diverging AI beliefs, lack of shared practices, and limited experimentation capacity, and organizational-level barriers, including uneven AI maturity, weak organizational learning mechanisms, and cross-unit coordination and alignment challenges. Our findings suggest that GenAI does not create new organizational challenges but becomes intertwined with existing tensions in large-scale agile transformation and may intensify challenges related to learning, coordination, governance, and trust. We contribute to large-scale agile transformation research by showing how GenAI interacts with ongoing organizational change and by identifying conditions for sustainable GenAI integration at scale.
\end{abstract}

\subsubsection*{Keywords:}

Large-scale Agile Development, Transformation, Autonomy, Generative AI, Teamwork

\section{Introduction}
\label{sec:intro}

Organizations undergoing large-scale agile transformations face a new challenge. While many have spent years establishing autonomous teams, product-oriented ways of working, and mechanisms for coordination and organizational learning, GenAI is rapidly becoming embedded in everyday work. Employees increasingly use AI through local experimentation and bottom-up adoption. While this may strengthen team autonomy and experimentation, it may also increase differences in practices and capabilities across teams, making organizational learning, coordination, and alignment more difficult. These are already persistent challenges in large-scale agile development \parencite{rath_systematic_2025,andersen2026tensions}.

GenAI is rapidly transforming software development and knowledge work. Tools such as ChatGPT, Claude, GitHub Copilot, and AI-enabled development environments are increasingly used to support coding, testing, documentation, analysis, and decision-making \parencite{ amalfitano_research_2026}. Recent advances suggest that AI is evolving from assisting individual tasks toward more autonomous forms of work, where multiple collaborating AI agents can decompose, coordinate, and execute development activities with limited human intervention \parencite{anthropic_agent_teams}. While much of the current discussion has focused on productivity gains and efficiency improvements at the individual level \parencite{Nguyen2023Generative,stackoverflow2024,Ebert2023,stray2025generative}, emerging evidence suggests that the implications of GenAI extend far beyond individual tasks. As GenAI becomes embedded in everyday work, it has the potential to reshape how agile teams collaborate, learn, coordinate, and make decisions \parencite{ amalfitano_research_2026,zhang_empowering_2025}.

While GenAI may enable faster experimentation, problem solving, and innovation \parencite{rajbhoj2024accelerating}, differences in how teams adopt and use GenAI may create challenges for governance, learning, coordination, and alignment \parencite{moe_stray2026aiagile}. Yet little is known about what happens when GenAI is introduced into organizations that simultaneously undergo large-scale agile transformation. Existing research has largely focused on AI tools, productivity, and individual developers \parencite{Ebert2023,stray2025generative,amalfitano_research_2026}, leaving the organizational implications of GenAI in large-scale agile settings underexplored. This matters because uneven GenAI adoption may reinforce coordination, learning, and alignment challenges that large-scale agile transformations seek to address. Successful GenAI integration therefore depends not only on access to technology, but also on organizational conditions that enable teams to experiment and learn while remaining aligned across organizational boundaries. We ask:

\textit{RQ: What barriers hinder the integration of generative AI in organizations undergoing large-scale agile transformation?}

To answer this question, we conducted a case study of TV 2, a large Nordic media organization currently undergoing a multi-year large-scale agile transformation. Because AI is being introduced while the organization continues to adapt its structures, coordination mechanisms, and ways of working, the case provides a unique opportunity to examine how AI interacts with ongoing organizational change.

\section{Background}
\label{sec:back}
\subsection{Large-Scale Agile Transformation}

The adoption of agile methods has expanded far beyond individual software teams and projects, leading to the emergence of large-scale agile development as a distinct research field \parencite{palopak_knowledge_2023,rath_systematic_2025}. Challenges related to scaling agile across multiple teams and organizational units have been discussed for more than a decade \parencite{xp2013}, and remain highly relevant as organizations continue to balance autonomy, coordination, and organizational learning at scale.

As organizations grow, the need for inter-team coordination increases, while teams must remain responsive to changing customer and market needs. %
Organizations therefore need to balance team autonomy with organizational alignment, support local adaptation while maintaining common ways of working, and coordinate decisions across multiple teams and organizational units \parencite{edison_comparing_2022,andersen2026tensions}. These challenges have led to the development of scaling frameworks such as SAFe and LeSS. However, research suggests that scaling depends not only on adopting a framework, but also on tailoring practices to the context and evolving coordination needs  \parencite{carroll_transformation_2023,edison_comparing_2022}. %

Recent studies emphasize that large-scale agile development should be understood as a continuous process of learning, adaptation, and coordination rather than the implementation of a predefined framework \parencite{dingsoyr_longitudinal_2022,uludag_revealing_2022}. These characteristics make large-scale agile organizations particularly interesting settings for studying the organizational implications of emerging technologies such as GenAI.

\subsection{GenAI in Large-Scale Agile} %

The adoption of GenAI tools is accelerating rapidly across software development. In the 2024 Stack Overflow developer survey, 76\% of 60,907 respondents reported that they were either using or planning to use AI tools in development work \parencite{stackoverflow2024}. A key driver of this adoption is the expectation that AI can increase productivity, improve software quality, and reduce effort across development activities.

Recent research suggests that GenAI may transform software development by supporting activities throughout the software life cycle \parencite{amalfitano_research_2026}. AI tools assist with code generation, testing, debugging, documentation, knowledge retrieval, and prototyping \parencite{Ebert2023,stray2025generative,rajbhoj2024accelerating}. While much of the existing literature focuses on productivity gains at the individual level, emerging research suggests that AI also affects how teams coordinate, learn, share knowledge, and make decisions \parencite{moe_stray2026aiagile,zhang_empowering_2025}. These implications become particularly significant in large-scale agile organizations, where organizational performance depends on how effectively autonomous teams coordinate, learn, and adapt together across organizational boundaries \parencite{dingsoyr_longitudinal_2022,palopak_knowledge_2023,rath_systematic_2025}. %

GenAI may strengthen these capabilities by enabling faster experimentation, prototyping, and problem solving. At the same time, it may create challenges related to governance, quality assurance, knowledge sharing, and coordination as teams adopt different tools and practices \parencite{amalfitano_research_2026}. Although previous studies have documented the productivity benefits of AI-assisted software development, less is known about how GenAI becomes integrated into everyday work practices in large-scale agile organizations. In particular, limited research has examined the barriers to AI-enabled learning, experimentation, coordination, and alignment across teams and organizational units. Successful adoption therefore depends not only on access to technology, but also on organizational mechanisms that enable teams to adapt while remaining aligned with broader organizational objectives.

\section{Method}
\label{sec:method}

To investigate our research question, we conducted an in-depth exploratory case study at TV 2, a Nordic media company undergoing a long-term large-scale agile transformation while simultaneously integrating generative AI into everyday work practices. An exploratory case study design was appropriate because it enabled us to examine how GenAI is understood, implemented, and managed within a complex organizational setting \parencite{yin2018case}.

\subsection{Large-scale agile transformation in TV 2}

TV 2 is a large Nordic media organization with approximately 1,000 employees, roughly equally distributed between Bergen and Oslo. The organization is undergoing a product-oriented transformation, with the technology unit currently comprising five product domains and approximately 15 product development teams of 5--10 members. Traditionally organized around television broadcasting and production, TV 2 has gradually shifted toward cross-functional teams, distributed decision-making, and new coordination mechanisms.

The transformation was initially driven by the technology department, which was reorganized into product domains serving different business and editorial functions. New roles were introduced to support coordination across units, including product managers, domain managers, and business representatives. The organization also adopted Objectives and Key Results (OKRs) to strengthen strategic alignment. Previous research conducted during the transformation, prior to the widespread adoption of GenAI, identified tensions related to work practices, role ambiguity, and cross-unit prioritization and decision-making \parencite{berntzen2026emerging}.

At the time of this study, agile structures had been established across significant parts of the organization, although the transformation remained ongoing. Simultaneously, GenAI was being adopted through local experimentation and bottom-up initiatives across technology, editorial, production, and support functions. TV 2 therefore provides a particularly relevant setting for examining how AI adoption interacts with ongoing challenges related to autonomy, coordination, learning, and alignment in large-scale agile organizations.

\subsection{Data Collection}
To obtain broad organizational insights and detailed accounts of AI use in practice, we combined two complementary data sources: semi-structured interviews and organizational survey data.

The qualitative data consists of 17 semi-structured interviews conducted between January and March 2026. To ensure variation in perspectives and experiences, we recruited participants at different organizational levels and units across the company (Table \ref{tab:participants}). Each interview lasted from 45 to 60 minutes. The interviews were conducted in pairs by the researchers, which allowed for multiple analytical perspectives during the conversations. The interviews followed a semi-structured interview guide while leaving room to pursue emerging topics, such as cultural tensions and concrete use cases of AI in everyday work. The interview guide is provided online: https://osf.io/zuw9p/files/23yzc.

The sample included technology managers and project managers in digital units working with core functionality and platform development, production managers responsible for the technical operation of studios, control rooms, and automated subtitling systems, as well as operational managers in the customer service unit responsible for AI-enabled chatbots and customer interaction. Table \ref{tab:participants} provides an overview of the respondents. To protect participant anonymity, respondents are described using broad organizational categories rather than specific job titles or departments. Organizational areas were grouped into editorial/content, production/operations, product/technology, and corporate/support functions (including HR, finance, marketing, customer support and legal). Managerial level was categorized as team-level, mid-level, or executive-level.%

\begin{table}[t]
\centering
\caption{Anonymized overview of participants}
\label{tab:participants}
\footnotesize
\begin{tabularx}{\columnwidth}{l X}
\toprule
\textbf{ID} & \textbf{Organizational area} \\
\midrule
P1  & Editorial/content \\
P2  & Production / operations \\
P3  & Editorial/content \\
P4  & Product / technology \\
P5  & Product / technology \\
P6  & Product / technology \\
P7  & Corporate / support functions \\
P8  & Product / technology \\
P9  & Production / operations \\
P10 & Corporate / support functions \\
P11 & Production / operations \\
P12 & Corporate / support functions \\
P13 & Corporate / support functions \\
P14 & Editorial/content/production \\
P15 & Corporate / support functions \\
P16 & Editorial/content \\
P17 & Product / technology \\
\bottomrule
\end{tabularx}
\end{table}

To complement the interview data, we were granted access to an internal organizational survey on AI use. The survey captured employees’ experiences with AI tools, perceived opportunities, challenges, concerns, and usage patterns across organizational units. While the survey was not designed as part of this research project, it provided an additional source of organizational-level data that allowed us to triangulate and contextualize findings emerging from the interviews.

Finally, we examined one software team identified internally as a particularly informative example of barriers to AI adoption (see illustrative example in Section 4). This provided a team-level perspective complementing the broader organizational findings.

\subsection{Qualitative Data Analysis}
All interviews were audio-recorded and transcribed automatically using OpenAI Whisper V3. To strengthen data quality, all transcripts were subsequently reviewed and manually verified by the research team. Names and other identifying or sensitive information were anonymized before analysis.

The analysis was conducted as an iterative process following the steps outlined by \parencite{pan_demystifying_2011}. Following the first rounds of interviews, the research team identified preliminary findings and illustrative quotations, which were organized around themes such as organizational culture, learning, experimentation, collaboration, and managerial responsibility. The process was conducted as follows: Each researcher were given a set of interviews for first-round coding. For each interview, we discussed and compared the codes and themes developed, resolving any disagreements by discussion. The analysis proceeded through repeated movement between close reading of individual interviews and comparison across interviews and data sources to identify recurring patterns, differences, and tensions. Through iterative comparison and refinement, these themes were consolidated into two overarching categories of barriers—team-level and organizational-level barriers—and their respective sub-themes, as reported in Sections 4.2 and 4.3. The themes were developed inductively and subsequently interpreted in relation to established tensions in large-scale agile research.

To strengthen the credibility of the study, preliminary findings were presented and discussed with participants from the organization. These discussions provided opportunities to validate, refine, and elaborate the researchers' interpretations and helped ensure that the analysis remained grounded in participants' experiences and perspectives.

\subsection{Quantitative Data Analysis}
The survey was administered internally at TV 2 in October 2025 and yielded 260 responses from employees and leaders across the organization's main units. %
Of the 260 respondents, 47 held leadership roles. The questionnaire combined closed questions on AI use, usage frequency, self-rated AI competence, perceived value, time saved, training needs, and the tools respondents used, together with a set of statements on workplace conditions and four open-text questions. Because the survey was not designed for this study, we treated it as a secondary source and used it to triangulate and contextualize the interview findings. %

We first prepared an analysis-ready dataset. We removed direct identifiers, normalized whitespace and blank-string artifacts, and derived a small set of indicators from the raw responses, including whether a respondent used AI at work, whether they used it daily, and whether they reported high or very high value from AI. Ordinal items, such as self-rated competence and the workplace-condition statements, were kept on their original scales. We conducted the analyses in Python and Excel. The analyses were descriptive and serve only to complement the interviews by showing how widely the patterns identified in the qualitative analysis were reflected across the wider organization. No causal claims are made. Percentages are based either on all 260 respondents or on the 213 who used AI, depending on who got the question, and every unit row in Table~\ref{tab:s2-unit-maturity} covers all 260.%

\begin{figure*}[tb]
\begin{illustrativecasebox}
\textbf{Illustrative Example: When AI Adoption Stalls}

One software development team, primarily consisting of backend developers, provided a particularly rich example of how diverging AI beliefs can undermine experimentation and collective learning. In an attempt to support responsible AI use, the team developed its own local AI guidelines. However, team members remained divided regarding how AI should be used and what constituted acceptable practice. One developer emerged as a strong AI enthusiast and began integrating AI throughout the development workflow, while another remained highly skeptical and preferred more traditional development practices. The disagreement intensified after a production issue was linked to AI-generated code.

After this incident, the AI-skeptical developer began scrutinizing the AI enthusiast's work more closely, while other developers became increasingly reluctant to review code they suspected had been generated using AI. According to the interviewee, discussions about AI became increasingly polarized and collaboration gradually declined. The  AI enthusiast, who had previously acted as an informal champion for experimentation, eventually withdrew from discussions and stopped driving AI-related initiatives.

Following these incidents, psychological safety deteriorated, and so did the team's willingness to experiment. Although team members agreed that AI would become increasingly important and identified several proof-of-concept initiatives worth pursuing, none were implemented. The team was unable to establish a shared direction for experimentation and failed to transform individual AI use into collective learning.

This case illustrates how AI adoption may stall even in teams that actively seek to explore new technologies. Rather than enabling continuous improvement and experimentation, AI became a source of disagreement that weakened collaboration, reduced psychological safety, and ultimately prevented the team from adapting its development practices.
\end{illustrativecasebox}
\end{figure*}

\section{Results}
\label{sec:results}

\subsection{Growing Interest, but Uneven Adoption}

Across interviews and survey responses, participants described GenAI as an important and increasingly unavoidable technology. Many interviewees expressed curiosity about AI and a desire to understand how it could support their work. Several participants also reported experimenting with GenAI tools, particularly within software development, content production, and analytical work.  Some participants described AI as a transformative capability that dramatically improved productivity: \textit{"Suddenly you've done a lot in a very short time. It is like having lots of consultants working for you."}[I04]. %
Others emphasized that AI had not yet become fully integrated into their everyday work practices. "\textit{We could have been a lot better at using AI, not least because we have spent a lot of time on it. So I think we have a lot of potential there}" [I15].
Others again struggled to identify meaningful use cases: \textit{"I do not really have a job where I have much use for it."} [I11]

At the same time, as can be seen in Table \ref{tab:s2-unit-maturity} the data revealed substantial variation in AI use, understanding, and maturity across the organization. Perhaps unsurprisingly, AI was found most valuable in the Product/Technology unit, where around half the respondents (46,9\%) used AI daily, and 70,8 \% found AI highly valuable. Moderate levels of use and value were found in the Finance, Production and Marketing units. Interestingly, AI use varied across the editorial units. In editorial A, a high number of respondents used AI (91.7\%)  and more than half of these reported daily AI use (55.6\%) and high perceived value of AI tools (63.6\%). In editorial unit B, on the other hand, only 58.8 \% used AI (and only 17.6 \% used it daily) and only 20\% of those users reported high value. This difference was attributed to the type of editorial content developed, but also differences in work cultures across the two units.

Overall, the interviews and survey data suggest that TV 2 is characterized by strong interest in AI and ongoing experimentation, but also by considerable differences in experience, competence, and adoption across the organization. The following sections present the barriers that participants identified as limiting broader and more coordinated AI adoption.

\begin{table*}[t]
\centering
\caption{AI adoption and maturity across organizational units ($N=260$). High value is the share of AI users in the unit reporting high or very high value from AI.}
\label{tab:s2-unit-maturity}
\setlength{\tabcolsep}{10pt}
\renewcommand{\arraystretch}{1.12}
\begin{tabular}{lrccc}
\hline
\textbf{Unit} & \textbf{$n$} & \textbf{Use AI (\%)} & \textbf{Daily use (\%)} & \textbf{High value among users (\%)} \\
\hline
Editorial A& 36  & 91.7 & 55.6 & 63.6 \\
Org./comm. \& legal            & 12  & 91.7 & 33.3 & 45.5 \\
Product/technology& 98  & 90.8 & 46.9 & 70.8 \\
Content strategy \& publishing & 17  & 82.4 & 23.5 & 64.3 \\
Finance                        & 14  & 78.6 & 28.6 & 45.5 \\
Production                     & 48  & 68.8 & 29.2 & 45.5 \\
Marketing                      & 18  & 66.7 & 16.7 & 41.7 \\
Editorial B& 17  & 58.8 & 17.6 & 20.0 \\
\hline
All respondents                & 260 & 81.9 & 37.7 & 58.7 \\
\hline
\end{tabular}
\end{table*}

\subsection{Team-level Barriers}

Participants identified several barriers at the team level, including diverging beliefs about AI, lack of shared AI practices, capacity constraints, and concerns surrounding trust, identity and psychological safety. 

\subsubsection{Diverging AI Beliefs Undermine Team Learning and Experimentation}

A recurring theme across the interviews was the presence of fundamentally different beliefs about AI. While some participants viewed AI as a transformative technology capable of significantly improving productivity and enabling new ways of working, others remained skeptical and questioned both its usefulness and reliability.  One participant described the current atmosphere around AI as positive: \textit{``The mood around AI is positive right now. It makes the working day easier. You have these moments of realizing, `I could just feed this into ChatGPT or Perplexity,' and suddenly you have saved seven hours.''} [P01]. Others, however, remained more skeptical and raised concerns about AI's future implications. As one participant reflected: ``\textit{But we both get really dystopian about it, and it's actually pretty bleak, a lot of what is emerging now. The last thing we said to each other was that before long neither of us will have a job anymore, and everything will have been taken over by AI}\ldots'' [I03]

The interviews suggest that AI adoption was often influenced more by informal opinion leaders than by formal organizational guidance. Strong individual opinions shaped the willingness to experiment and discuss AI openly. This dynamic is illustrated in the example on the previous page%
, where disagreement between an AI enthusiast and a skeptic gradually reduced trust, weakened collaboration, and prevented the team from pursuing AI-related experimentation.

\subsubsection{Lack of Shared AI Practices Inhibits Collective Experimentation}

Although AI tools were widely available, participants repeatedly described AI adoption as highly individual and largely uncoordinated. Employees experimented with different tools, prompting strategies, and workflows, but few teams had established shared practices regarding when and how AI should be used. One interviewee described the organization's culture as highly autonomous: \textit{"It is part of the TV 2 culture that anything goes. You can experiment and adopt whatever you want."} [P08] %

While this autonomy encouraged experimentation, it also resulted in fragmented practices. Teams and individuals often developed their own solutions, used different tools, and rarely discussed AI use systematically. As one HR leader observed: \textit{"I probably should spend more time with [team member who leads AI agent initiatives], she is a bit alone in this. We should spend more time together at workshops and seminars to find inspiration for how to do things differently."} [I7]

Several participants emphasized that learning occurred primarily through individual exploration rather than collective reflection. As a result, successful practices remained local and were rarely scaled beyond individual teams and departments. %
The lack of a collective approach towards AI adoption also led to what some interviewees in certain departments pointed to as a "generational gap", where elderly employees were frequently slow to test out new AI-based tools without getting any assistance. I13 said: \textit{"There is a very big difference depending on age. My employees, those who use [AI] and those who think it is very scary. There are very different levels."}

\subsubsection{Limited Capacity Constrains Continuous Improvement}

Many participants highlighted the lack of time available for experimentation and learning. Despite recognizing the potential value of AI, employees often prioritized operational responsibilities and ongoing delivery commitments, which could mean they did not get to practice their AI skills: \textit{For example, I remember when I took that course myself, the part about Perplexity — I just thought, wow! And then I haven’t really used it for a few months now. So it’s important to keep at it, kind of.} [I01]%

Several interviewees argued that teams needed dedicated time to explore new ways of working and exchange experiences: \textit{"We could probably benefit from spending more time together to explore possibilities for AI use."}[I07] %

These findings suggest that AI adoption requires more than access to technology. It also requires organizational conditions that allow teams to engage in continuous learning and experimentation.

\subsubsection{Trust, Psychological Safety, and Professional Identity Concerns}

The interviews revealed concerns regarding trust, quality, and professional identity across multiple organizational units. %

Several participants questioned whether AI-generated outputs could be trusted in contexts where accuracy and credibility were critical. Journalists emphasized the importance of maintaining editorial standards and public trust: \textit{"You can never compromise credibility and facts."} [I11] %
Another editorial employee expressed similar concerns: \textit{"It is my name, my signature, and my identity that are attached to a piece written by a robot."} [I01] %
Trusting AI outputs could also negatively affect trust between team members, which P08 illustrated by pointing to how he had been in situations where he had to call out team members for sloppy AI use: \textit{"Some people on the team uncritically rely on answers from an AI. You can get all sorts of things from it, but that's a problem for us, given that we know LLMs can still hallucinate quite a lot."}

Others expressed concern that AI-generated content might weaken the personal and human aspects of work: \textit{"You should meet people with presence, and you should meet them in text and form in a report."} [P07] %

The illustrative software team example above %
further demonstrated how lack of trust and psychological safety can influence AI adoption.

\subsection{Organizational-level Barriers}
At the organizational level, barriers included fragmented governance, uneven cross-unit maturity, underdeveloped organizational learning mechanisms, autonomy-alignment tensions, and cross-unit collaboration challenges during AI adoption. 

\subsubsection{Fragmented Governance and Lack of Alignment}

A major organizational barrier concerned the absence of a shared direction for AI adoption. Participants repeatedly described uncertainty regarding expectations, priorities, and acceptable practices.

Although general guidelines existed, employees often interpreted them differently. Several participants called for clearer organizational guidance and more consistent communication regarding AI strategy.

The absence of shared direction resulted in local experimentation but limited organizational alignment. Teams adopted different tools, developed local practices, and pursued AI initiatives independently. It was evident through the interviews that the views on what was acceptable use of AI also differed between the organizational units. One head of content development [I3] who had been tasked with attempting to create common understandings of AI use across units recalled this scenario: \textit{"When we were giving a pitch presentation of [a popular TV show], an original image of a location that doesn't exist was included, showing a house on the edge of a forest. We weren't able to find it [as it was AI generated]. And when something puts images in our heads that are not fulfilled later, it can create feelings of being deceived}."

\subsubsection{Uneven AI Maturity Across Organizational Units}

As described in Section 4.1, both the survey and interviews revealed substantial differences in AI adoption and maturity across organizational units. %
Table \ref{tab:s2-unit-maturity} shows particularly high AI use and perceived value in Product/Technology and one editorial unit, while adoption was considerably lower in other parts of the organization. These differences were also reflected in the interviews. One participant explained: \textit{"I think when it comes to editorial staff, this may be one of those professional groups that embraces AI. Maybe it has to do with the profession, and with being curious and creative, and always interested in what’s new. So it may be different than other professional groups"} [I01]. %

A respondent from a corporate/support function described it as follows: "\textit{We are not the most advanced in using AI, but we do have a number of communities that are quite good at experimenting and testing it.}" [I15] %
These differences created challenges for coordination and knowledge sharing, as teams and units progressed at different speeds and developed different understandings of what AI could and should be used for.

\subsubsection{Underdeveloped Organizational Learning Mechanisms}

While individual experimentation was common, participants frequently described organizational learning mechanisms as fragmented and insufficiently developed. Specifically, the interviewees requested more opportunities to share experiences, discuss failures, and learn from successful implementations. One participant argued that employees needed more examples of what could go wrong rather than more examples of successful use: \textit{"We do not need more best practices. We need worst-case examples."} [I03] %
Others highlighted the need for communities of practice, AI champions, and cross-team learning arenas. One leader in a creative department [I16] called for more AI champions across the organization, not just connected to the technology department, that could share tried-and-tested practices for the specific communities: \textit{"It would be good if someone could make it a bit more tangible for people who aren't very familiar with coding. [...] You get confirmation that you can actually use it without having to think, 'It's going to take at least two years before I'm able to do this.'"}  In other words, without targeted mechanisms for knowledge sharing, learning remained localized and difficult to scale.

Participants described GenAI adoption as unfolding alongside—and competing for attention with—the ongoing large-scale agile transformation. While the organization continued its efforts to strengthen product-oriented ways of working, cross-unit collaboration, and organizational learning, participants observed that AI-related initiatives increasingly attracted attention and resources. Some expressed concern that maintaining momentum across multiple transformation efforts simultaneously could become challenging. This suggests that AI adoption may not occur independently of other organizational change initiatives, but becomes intertwined with ongoing transformation processes.

\subsubsection{Autonomy–Alignment Tensions}

A central element of TV 2’s large-scale agile transformation was increased team autonomy and distributed decision-making. While this supported local GenAI experimentation, it also created challenges when autonomous teams adopted AI in different ways. Several participants described how teams selected different tools, developed different practices, and interpreted organizational guidance differently. Thus, GenAI adoption became intertwined with an existing challenge in the agile transformation, namely balancing local autonomy with organizational alignment:
"\textit{We cannot force people to use these tools, but we can try to facilitate their use.}" [I12] %

\subsubsection{Cross-Unit Coordination Challenges}

Finally, cross-unit coordination, already a central challenge in TV 2's large-scale agile transformation, also shaped the organization's ability to coordinate GenAI adoption. Different units had different responsibilities, work practices, and risk profiles. Editorial environments emphasized credibility and trust, technology teams focused on productivity and development efficiency, while operational units prioritized reliability and stability. These differences complicated efforts to establish common approaches to AI adoption. As AI use expanded, coordination across teams and organizational units became increasingly important, yet many participants perceived existing coordination mechanisms as insufficient. Not only was this due to disciplinary differences, but also physical location. One participant was concerned about how the lack of physical meetings affected a common organizational culture needed for a united approach to AI-led transformation: \textit{"We are probably a bit too separate as departments, and we would have liked to be able to say that we worked together on things more often. There is, overall, a bit too much of a cultural divide between the different locations. [...] And if a relationship mostly consists of a short Teams meeting or a few Slack messages, most of the time, then it is easier to be misunderstood."}[I14]

\section{Discussion}
\label{sec:discussion}
\subsection{Growing Interest and Experimentation}

Our findings indicate that TV 2 possessed several conditions that enabled GenAI adoption. Across the organization, participants described strong interest in GenAI, widespread experimentation, and a general belief that AI would become increasingly important for future work practices. Experimentation was observed across software development, content production, analytics, customer service, and administrative functions, suggesting that AI adoption was not confined to a single professional domain.

The findings further suggest that the organization's large-scale agile structures supported local experimentation. Team autonomy and openness to new technology enabled employees to explore new tools and ways of working without waiting for centrally coordinated initiatives. At the same time, participants emphasized the need for stronger mechanisms for knowledge sharing and organizational learning.

Taken together, our findings suggest that the challenge was not a lack of interest in AI, but rather the organization's ability to transform local experimentation into coordinated organizational learning and adoption. %

\subsection{Barriers to GenAI Integration in Large-Scale Agile Organizations}
To answer to our research question, we identified barriers to GenAI integration at both the team and organizational levels. At the team level, barriers were associated with diverging beliefs about GenAI, lack of shared practices, limited opportunities for experimentation, and concerns related to trust, quality, and professional identity. At the organizational level, barriers were linked to fragmented governance, uneven AI maturity across organizational units, limited mechanisms for organizational learning, and challenges related to coordination and alignment. Similar patterns have been observed in other media organizations, where GenAI use is widespread but highly individualized, with limited coordination and knowledge sharing across teams  \parencite{xiao2026collaborativenewswork}.

The findings related to uneven levels of AI maturity may reflect differences in digital mindsets, defined as attitudes and beliefs toward new technology \parencite{solberg_digital_2020}. The type of mindset may determine whether individuals believe in their ability to develop new skills and see AI as a tool for growth and development, or whether they become more reactive. As illustrated by the software team example, mindset differences may influence the extent to which teams are able to experiment and learn which in turn can affect the extent to which the organization is able to successfully leverage GenAI.

Taken together, the findings suggest that integrating AI into large-scale agile organizations is not primarily a technological challenge. Rather, successful integration depends on organizational mechanisms that enable teams to learn, experiment, coordinate, and develop shared ways of working across the organization \parencite{hansen_understanding_2024,zhang_empowering_2025}. This supports research arguing that AI adoption should be understood as a socio-technical change process rather than the deployment of a new tool \parencite{moe_stray2026aiagile}.

A central insight from this study is that AI does not appear to create entirely new organizational challenges. Previous research has highlighted persistent tensions between autonomy and alignment, local adaptation and organizational coordination, and experimentation and governance in large-scale agile organizations \parencite{rath_systematic_2025,andersen2026tensions}. Our findings suggest that the introduction of AI exposes and may intensify these existing tensions by increasing the speed of experimentation, decision-making, and local innovation. For example, AI enables individuals and teams to experiment more independently, but this simultaneously increases the need for coordination, governance, and shared practices across teams \parencite{amalfitano_research_2026}. Rather than introducing a fundamentally new set of organizational challenges, AI may reinforce tensions that large-scale agile organizations are already required to manage.

\subsection{Implications for Large-Scale Agile Transformation}
Our findings show how GenAI adoption became intertwined with several mechanisms and tensions already central to TV 2’s large-scale agile transformation. Rather than emerging as a separate technological initiative, AI adoption unfolded through many of the same organizational mechanisms that were already central to the ongoing transformation. %

This study contributes to research on large-scale agile transformation by extending our understanding of how emerging technologies interact with agile organizational structures. Existing research has focused on scaling frameworks, coordination, autonomy, alignment, and organizational design \parencite{edison_comparing_2022,andersen2026tensions}. Our findings suggest that AI is a new organizational force interacting with these mechanisms. What distinguishes GenAI in our case is rapid bottom-up experimentation, individualized adoption, uncertainty about acceptable use and output reliability, and differences in practices and maturity across teams and organizational units.

More specifically, the study highlights the importance of organizational learning mechanisms in AI-enabled large-scale environments. Agile organizations rely on experimentation, knowledge sharing, and continuous learning \parencite{rath_systematic_2025}, but our findings suggest that these mechanisms become even more important when AI is introduced. Without structures that support collective learning, organizations risk fragmented adoption, local optimization, and increasing differences in work practices across teams. %

Similarly, the team-level case demonstrated how AI adoption depended on mechanisms such as psychological safety, experimentation, and informal change agents—factors that have long been recognized as enablers of agile transformation and continuous improvement \parencite{dikert2016challenges, rath_systematic_2025}.

The findings further suggest that AI may reinforce one of the core tensions identified in large-scale agile research: the balance between team autonomy and organizational alignment \parencite{dikert2016challenges,ravn2022team}. AI tools increase the ability of individuals and teams to experiment, automate work, and develop local solutions. However, as teams adopt different tools, establish different practices, and 
progress at different speeds, the need for shared standards and cross-team coordination also increases. In this sense, AI does not eliminate existing tensions but may reinforce them.

Finally, our findings indicate that organizations may need to manage multiple transformation agendas simultaneously. TV 2 was already engaged in a long-term transformation focused on product-oriented ways of working, cross-unit collaboration, and organizational learning when AI adoption accelerated across the organization. As a result, AI emerged not only as a new technology, but also as a new source of organizational attention, experimentation, and change. This observation extends previous research on large-scale agile transformation, which has highlighted tensions between competing organizational priorities and transformation initiatives \parencite{andersen2026tensions}. Rather than representing a separate transformation, AI became intertwined with the ongoing agile transformation.

\section{Conclusion and Future Work}
\label{sec:conclusion}
This study examined GenAI integration in a large Nordic media organization undergoing a large-scale agile transformation. Our findings suggest that successful GenAI integration is an organizational rather than a technological challenge. Rather than creating new challenges, GenAI became intertwined with the ongoing transformation and with existing tensions related to autonomy, alignment, learning, and coordination.
As a single-case study, our findings are shaped by the organizational and industry context of TV 2. In particular, concerns related to editorial credibility, public trust, content authenticity, and professional identity may be especially salient in media organizations and may manifest differently in other software-intensive organizations. Future studies across different industries are therefore needed to examine the transferability of the identified barriers.
Future research should examine how organizations can support collective learning, coordination, and governance in AI-enabled large-scale agile environments. Longitudinal studies would be particularly valuable for understanding how AI adoption interacts with ongoing transformation efforts over time.

\section*{Acknowledgments}
We thank TV 2, the interview participants, and the survey respondents for their valuable contributions. 
This work was supported by the Research Council of Norway through the project KAIROS (grant 357147).

\printbibliography

\end{document}